# Effects of Geometry and Symmetry on Electron Transport through Graphene-Carbon-Chain Junctions

Yao-Jun Dong[1], Xue-Feng Wang[1,2*], Ming-Xing Zhai[1], Jian-Chun Wu[3], Liping Zhou[1], Qin Han[1], Xue-Mei Wu[1,2]

[1] Department of Physics, Soochow University, 1 Shizi Street, Suzhou 215006, China

[2] State Key Laboratory of Functional Materials for Informatics, Shanghai Institute of Microsystem and Information Technology, Chinese Academy of Sciences, 865 Changning Road, Shanghai 200050, China

[3] Key Laboratory for Radiation Physics and Technology of Education Ministry of China, Institute of Nuclear Science and Technology, Sichuan University, 29 Wangjiang Road, Chengdu, 610064, China

*E-mail: xf_wang1969@yahoo.com

**Abstract**

The electron transport between two zigzag graphene nanoribbons (ZGNRs) connected by carbon atomic chains has been investigated by the non-equilibrium Green's function method combined with the density functional theory. The symmetry of the orbitals in the carbon chain selects critically the modes and energies of the transporting electrons. The electron transport near the Fermi energy can be well-manipulated by the position and the number of carbon chains contacting the nanoribbons. In symmetric ZGNRs connected by a central carbon chain, a square conductance step appears at the Fermi energy because the antisymmetric modes below it are not allowed to go through the chain. These modes can additionally contribute to the conductance if side carbon chains are added in the connection. By choosing a proper geometry configuration, we can realize Ohmic contact, current stabilizer, or the negative differential resistance phenomenon in the devices.

**Introduction**

Because of the spectacular electronic properties of carbon-based materials, carbon-based nanostructures have attracted great attention for their possible application in electronics and spintronics[1-3]. The quasi-one-dimensional (Q1D) electron transport in carbon nanotubes (CNTs), discovered in 1991, have been intensively studied[4]. Graphene, which was mechanically exfoliated in 2004 by Geim[5], has remarkable mechanical, electrical, and thermal properties. The 2D graphene and the Q1D graphene nanoribbons (GNRs) have been applied in designing nanoelectronic devices for their special electronic transport properties such as the high mobility and the magnetic edge effects[6-8]. The field-effect-transistors (FETs) fabricated by carving a constriction in graphene[1] have attracted a massive interest in such geometries, and FETs based on GNRs have already been fabricated by Ponomarenko *et al*[6]. Because the downscale of FETs intended to be sub-10 nm, the semiconducting GNR-FETs is limited to the current lithography technique[7].

To scale down the system dimension, employing carbon-atomic-chains (CACs) is a promising choice[9, 10]. For decades, CACs were difficult to synthesize owing to the high reactivity of chain ends and a interchain cross-linking.[11] Theoretical research of CACs connected to metallic electrodes has been carried out since 1998. Lang *et al*.[12, 13] have shown that the conductance depends on the parity of the atom number in the chain as well as the contact between the chain and the metal electrodes. Larade *et al*.[14] have observed negative differential resistance (NDR) due to the competition between the contact effects and the contribution of density of states. Deng *et al*.[15, 16] have also predicted NDR in a CAC junction connected to gold electrodes. Tongay *et al*.[17] have concluded that the CAC structure is stable and can be easily doped, functionalized, or used to interconnect molecular devices. Crljen *et al*.[18] have found that the contact conductance to gold electrodes by CACs can be one order of magnitude larger than that by other conjugated oligomers and suggested that CACs may be used in nanoelectronics. Khoo *et al*.[19] have shown that the NDR occurs when CACs are connected to metallic CNT covalently. Although CACs were extensively interesting during the last decades, the synthesis of CACs connected to metal electrodes remains very difficult in practice.

Recently, CACs have been successfully fabricated from graphene using a high energy electron beam[20-24]. This advance is significant because CACs may be used to connect graphene-based devices[8, 25]. A few theoretical studies have been performed on the electron transport properties in graphene-carbon-chain junctions. Shen *et al.*[26] have observed the same length effect on conductance as in the case of metal electrode[12, 13] and found that the ballistic transport is not sensitive to the existence of disorders in long CACs. The structure has also been predicted a perfect spin filter and spin valve[27, 28]. One of the characteristics that intrigues us is that the conductance vanishes below the Fermi energy in some cases[27, 29, 30] while remaining finite in other cases[26, 31, 32]. Various possible connecting ways between graphene and carbon chain have been assumed in previous works and the symmetry selection effect on the connections may have led to the diversified results.

Contacts are necessarily involved in FETs devices and play a quite important role in ballistic transport[33, 34]. In the last years, many studies have addressed the contact effect in graphene-based transistors[35, 36]. It has been shown that the contact type and contact position may determine the $I_{ON}$ to $I_{OFF}$ ratio[37, 38] and contact resistance[39-41]. The height of the Schottky barrier at CNT-graphene contacts can be about one order lower than that at Pd-graphene contacts[38]. For Au-graphene contacts, the ratio of the end-contacted resistance per carbon atom to the side-contacted one can be as huge as 6000.[41] We will focus on the effect of geometry and symmetry on the charge transport in junctions of CACs and zigzag GNRs (ZGNRs), the promising materials for nanodevices.

**Computational Models and Methods**

There are three typical atomistic structures of a CAC-ZGNR contact point, namely the three-, five-, and six-membered rings[26, 27, 29, 30]. Our calculation indicates that the choice of the contact point structure do not affect the conclusion obtained from this work. Here we use the three-membered-ring structure connecting a CAC to an *n*-ZGNRs with even width number *n*=6.[42] There are five typical geometry configurations of connecting two aligned 6-ZGNRs by CACs parallel to the edges of the ZGNRs. Treating the 6-ZGNRs as electrodes, we establish five two-probe devices, D1-D5, as shown in Figures 1a and 1d -1g, respectively. Each device is partitioned into the left electrode (L), the central scattering region (C), and the right electrode (R). Because ballistic transport maintains in odd-numbered chains[12, 26] and we have confirmed that our conclusion is independent of the length of the chain, we adopt a seven-atom CACs in the simulation. On each end of the CACs in the scattering region, a buffer layer of 6-ZGNR is added to ensure that the contact effect on the electrostatic potential in the electrodes is negligible. Since the energy separation between the antiferromagnetic and ferromagnetic states in ZGNRs is small[43, 44] and the spin effect is not important at high temperature, we treat the systems as nonmagnetic. Note that the result may vary quantitatively in some cases if the spin effects are taken into account[45, 46].

Using the three-membered-ring structure, we can easily show the symmetry selection effect on electron transport at the contact area. The orbitals of π electrons transporting charges of the three C atoms in the contact area at each contact point are illustrated in Figures 1b-c. The $p_x$ ($p_y$) orbitals of the end atom in the CAC have symmetric parity about the $x$ ($y$) axis and antisymmetric parity with respect to the plane perpendicular to $x$ ($y$) axis. Only the $p_y$ orbitals of the two atoms in the 6-ZGNRs are involved in the charge transport. Here we define the $x$ direction as the width direction of the ZGNR. In this case, the electrons in the $p_x$ orbitals in the CACs are localized and the electrons in the $p_y$ orbitals in the CACs can penetrate extensively into the electrodes only if the bond between the $p_y$ orbitals of the two atoms in the 6-ZGNRs has symmetric (with respect to the CAC axis) component.

In the simulation process, first geometry optimization for each system is performed by the Vienna ab initio simulation package (VASP)[47, 48] and the atomistix toolkits (ATK) to adjust the structure parameters until the force felt by each atom becomes < 0.02 eV/Å. Second, the eletron-transport properties are calculated by the nonequilibrium Green's functions method (NEGF) and the *ab initio* density function theory (DFT) implemented in the ATK package[49, 50]. The exchange correlation potential resorts to the local density approximation with the Perdew-Zunger parametrization (LDA-PZ), which leads to the same conclusion as the generalized gradient approximation with the Perdew-Burke-Ernzerhof parametrization (GGA-PBE) in this work. The wave functions are expanded on a basis set of double-ζ orbitals plus one polarization orbital (DZP), which can preserve an accurate description of π bonds. A 15Å thick vacuum layer is adopted to separate the ribbons in both $x$ and $y$ directional neighbor supercells far away enough and to ensure the suppression of the coupling between them. In the calculation, the energy cutoff is 150 Ry, the mesh grid in the $k$-space is 1×1×100, and an electronic temperature of 300K is used in the technique of the eal-axis integration for the nonequilibrium Green's functions. To help analyze the result and understand the geometry effect in the junctions of the GNRs and the CACs, we construct virtual bulks to illustrate the pattern of the wave functions and their corresponding energies in the junctions. The virtual bulks are periodic along the $z$ axis with a unit cell enclosing the junction regions in each two-probe device as illustrated by the insets in Figure 2a. We calculate the energy bands and the corresponding wave functions of the virtual bulks to extract extra information about the junctions. By comparing the wave function of the virtual bulks with the local density of states (LDOS) of the two-probe devices at nearby energies, we can identify the transport modes and figure out their parity.

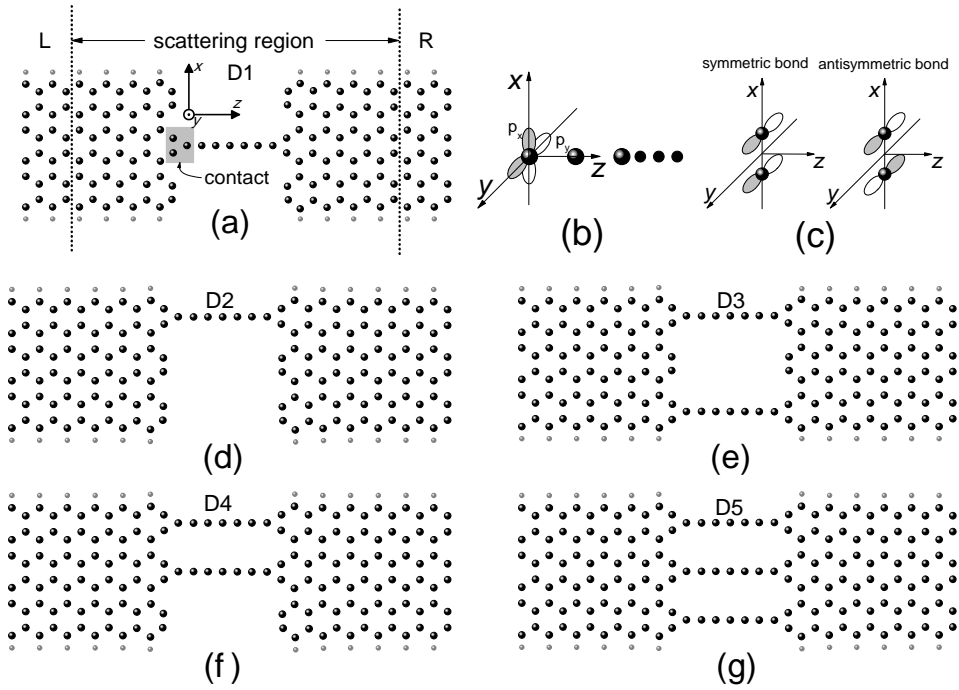

Figure 1. Top view of devices D1-D5 are shown in (a), (d), (e), (f), and (g), respectively. The left and right 6-ZGNR electrodes are connected by one or more CACs in each device. The big spheres indicate the positions of the C atoms and the small ones the H atoms passivating the dangling bonds on the edges. The shadow in (a) indicates the contact area between the left electrode and the CACs. The $p_x$ and $p_y$ orbitals of the C atom of CAC in the contact area are schemed in (b) and the $p_y$ orbitals of the two C atoms of electrode in the contact area are schemed in (c) for the symmetric bond and the antisymmetric bond. The electron transport is along the $z$ axis.

**Results and discussion**

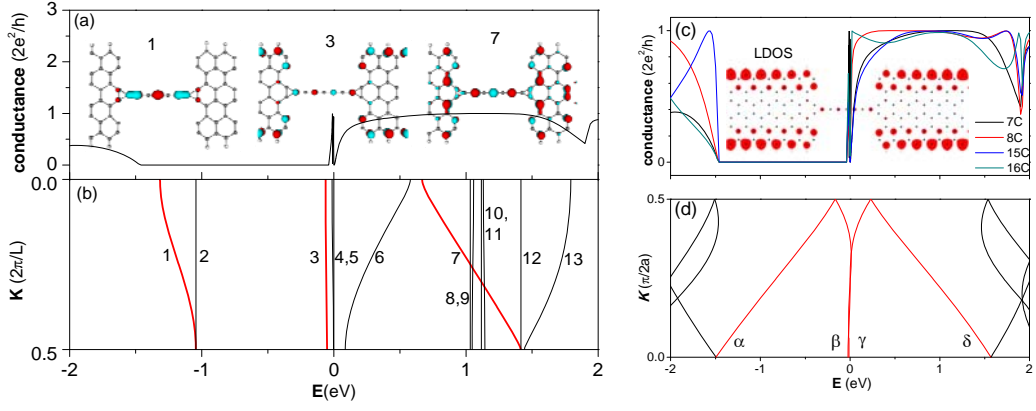

Figure 2. (a) Solid curve shows the conductance spectrum of device D1. (b) Energy bands in the Brillouin zone of the virtual bulk corresponding to device D1 where $L$ is the length of its unitcell. The 3D isosurface plots of the wave functions for k=0 states in bands 1, 3, and 7 are shown as insets in (a). (c) The conductance spectrum of device D1 with CACs of 7 (black), 8 (red), 15 (blue), and 16 (green) C atoms. The LDOS at E=0 in D1 with a seven-atom CAC is shown in the inset. (d) Energy bands of a 6-ZGNR bulk calculated with unitcell of four primitive cells. Here $a = \sqrt{3} a_{cc}$ with $a_{cc} = 1.42$ Å the C-C bond length in graphene. The three bond lengths in the 7-atom CAC are 1.354, 1.290, and 1.312 Å from the end to the center, respectively. The energy $E$ is reckoned from the Fermi energy $E_F$=0.

In Figure 2a, we present the conductance spectrum of device D1 where the left and the right 6-ZGNRs are connected by a seven-atom CAC along the centerline of the ribbons. The system has mirror symmetry with respect to the CAC. An almost step shape spectrum is observed with a sharp peak at $E_F$, a zero plateau below the Fermi energy $E_F$, and a one-conductance-quantum plateau above $E_F$. To figure out the transport modes near $E_F$, we plot the energy bands of the virtual bulk corresponding to D1 in Figure 2b and present some of the wave functions as insets of Figure 2a to show the corresponding transport modes. Below $E_F$, there are five energy bands. The wave function of the states in band 1 (transport mode 1) is mainly confined in the chain with symmetric parity. This mode cannot transport charge in the two-probe device D1, although its energy band has dispersion and shows some extending characteristics in the virtual bulk. This indicates that its localization length is longer than the unit cell length of the virtual bulk but this mode cannot enter the 6-ZGNR electrodes. It is a result of symmetry selection. As shown in Figure 2d, the wave function of states (α and β) near and below $E_F$ in 6-ZGNRs has antisymmetric parity and the bond between the $p_y$ orbitals of the two atoms of the electrodes in the contact point has no symmetric component. As a result the electrons in transport mode 1 (localized in the CAC) and mode 2 (localized in the electrodes) cannot flow freely in the two-probe device D1. The wave function of the states in band 3 (transport mode 3) has symmetric parity and can transport charge. It is confined to the edges in the ZGNRs so corresponds to the transport mode of the edge states at $E_F$ in the two-probe device D1. This is confirmed by the calculation of the local density of states (LDOS) at $E_F$ in the two-probe device D1 as shown in the inset of Figure 1c. In the virtual bulk system, a charge transfer occurs between the CAC and the ZGNR parts and $E_F$ shifts slightly to a higher energy compared with that in the two-probe device. The states in bands 4 and 5 are composed of orbitals in the CAC and are localized. The states in bands 6, 7, and 13 are all symmetric in both CAC and ZGNRs but have different patterns. These modes can carry free electrons and result in the one-conductance-quantum plateau. The states in bands 8, 9, 10, and 11 correspond to the dangling orbitals at the ends of the electrodes and are localized. The states in band 12 are localized in the ZGNRs with antisymmetric parity and come from higher bands of the ZGNRs shown in Figure 2d.

The conductance spectra of device D1 with CAC of different lengths are plotted in Figure 2c. All of them have step shape structure at $E_F$. A wide $G=0$ conductance gap exists in the energy range [−1.47, 0] eV followed by a wide $G=2e^2/h$ conductance plateau in the range [0, 2] eV with some sharp peaks and dips near $E_F$ and 1.9eV. As shown in Figure 2d, wavefunction of states in the energy range [-1.47, 0] eV (bands 1 and 2 in red) in 6-ZGNRs have antisymmetric parity and electrons in those states cannot pass through the CAC. At energy less than -1.47eV, states with symmetric components emerge and the conductance increases. Above $E_F$ there is one state with symmetry parity and the conductance can reach to the conductance quantum. The dips at the end of the plateau are result of the interference between the states in the CAC and the ZGNR.

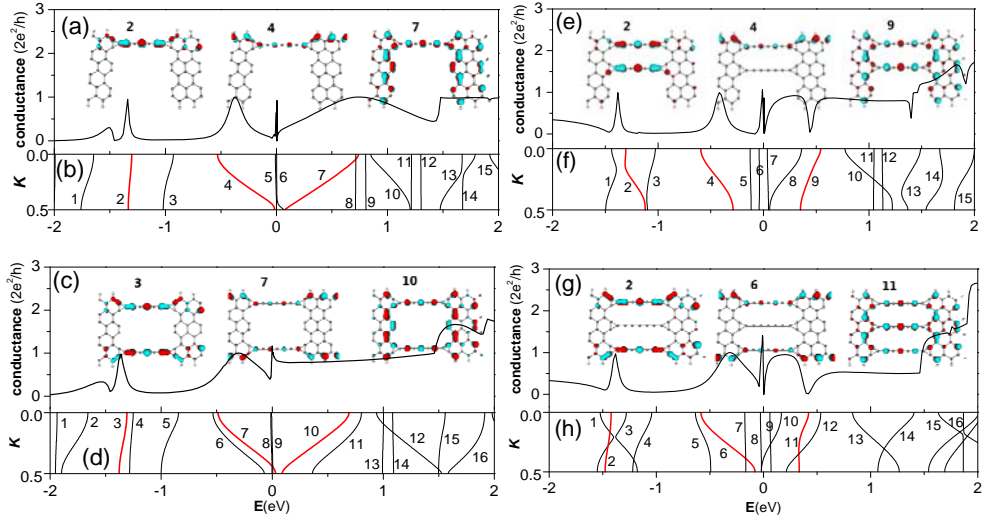

Figure 3. Conductance spectra of devices D2, D3, D4, and D5 are plotted in (a), (c), (e), and (g), respectively. Some wave functions of the virtual bulk corresponding to each device are illustrated as insets to show the transport modes. The energy bands of the virtual bulks corresponding to devices D2, D3, D4, and D5 are presented in (b), (d), (f), and (h), respectively. The three bond lengths of the seven-atom CAC in the side position are 1.360, 1.289, and 1.314 Å from the end to the center, respectively.

In device D2, the two ZGNR electrodes are connected by one CAC on one side. The mirror symmetry with respect to the CAC shown in D1 is removed. For any incident electrons from the ZGNR electrodes, with symmetric or antisymmetric parity, there is usually a symmetric component in the bond between the $p_y$ orbitals of the two ribbon atoms in the contact area (see Figure 1b). The conductance then become finite below $E_F$ and mainly two conductance peaks (at -1.3 and -0.3 eV) appear as shown in Figure 3a. With the help of the electronic energy bands and the wave functions in the virtual bulk of D2 as shown in Figure 3b, we obtain the corresponding transport modes. The peak at -1.3 eV comes from a transport mode where electrons flow along the chain side of the 6-ZGNR and the peak at -0.3eV is a result of the edge states in the ribbon because it is closer to $E_F$. Above $E_F$, there is a wide conductance peak and the corresponding transport mode 7 looks like mode 7 of D1 in the ZGNR region. At k=0, band 5 corresponds to a mode confined in the CAC and band 6 corresponds to a mode confined to the lower side of the ZGNRs. At $E_F$, mode 4 mixes with mode 6 and band 4 splits near k=0.5. The LDOS at $E_F$ appears a pattern composed of more than modes and indicates that the conductance peak originates mainly from transport mode 4.

If we add another CAC on the other side of D2, we have a symmetric device again (D3) with respect to the central line of the electrodes. As shown in Figure 3c, the conductance spectrum of device D3 has peaks in the same energies as that of D2. Those peaks becomes wider because the capacity of the transport channels is enhanced by the added CAC. In the band structure of the virtual bulk for D3 as shown in Figure 3d, some of the energy bands corresponding to states in the CACs double and split due to the existence of double chains. The energy bands 2, 4, and 7 of the virtual bulk for D2 shown in Figure 3b split into bands 3/4, 6/7, and 10/11 of the virtual bulk for D3, respectively. The transport modes in D3, illustrated as insets in Figure 3c, can be derived from those in D2 by adding an antisymmetric counterpart on the side of the added CAC for modes below $E_F$ and by adding a symmetric one for modes above $E_F$. The conductance peak at $E_F$ shows characteristics of Fano resonance due to the weak coupling between the extending modes 7 and 10 and the localized modes 8 and 9.

Adding a CAC on one side of D1, we have device D4 as shown in Figure 1f. For $E < E_F$, two same conductance peaks as in D2 emerge from zero due to the added transport channels through the side CAC. For $E > E_F$, the conductance decreases and two antiresonance dips appear at 0.45 eV and 1.3 eV due to the interference between transport modes through the two CACs as illustrated in Figure 3e. Three transport modes (2, 4 and 9) having the same energies as the two peaks and the low-energy dip, respectively, are plotted as the insets of Figure 3e and the energy bands of the virtual bulk corresponding to device D4 are presented in Figure 3f.

When we add another CAC on the lower side of D4, we obtain the symmetric device D5 with triple CACs. As shown in Figure 3g, the conductance spectrum of D5 has peaks and dips at the same energies as that of D4. For $E < E_F$, the peaks become wider and have the same shape as those of D3. The energy bands and wavefunctions in the corresponding virtual bulk as presented in Figures 3g and 3h indicate that the typical transport modes in D5 are the same as those in D3. The central CAC does not participate in the transport for the same reason as in D1. For $E > E_F$, the dips become wider and the conductance decreases overall showing stronger anti-resonance between the central and side CACs. The peak at $E = E_F$ becomes much higher than that in other devices with a maxima value near $1.5 \times 2e^2/h$. This indicates that the electrons in the edge states of ZGNRs can pass almost freely through the triple CACs.

From the previous discussion, we see that the geometry symmetry plays a key role in determining the conductance spectra of junctions of CACs connecting two transversally symmetric electrodes of 6-ZGNR. When only one CAC connects the electrodes along their centerline, the spectrum has a step form because the electrons in the antisymmetric states below the Fermi energy cannot pass through the CAC. When one CAC connects the electrodes on one side, the conductance is determined by the symmetry of the wave function in the contact region. Resonant sharp conductance peaks appear below the Fermi energy and wide peaks above the Fermi energy. In devices with one centerline CAC and one side CAC, the interference between the two CACs introduce antiresonant conductance dips above the Fermi energy while the sharp peaks below the Fermi energy corresponding to the side CAC remain. When two side CACs exists, the additive characteristics appear where the sharp resonant peaks and antiresonant dips widen as a result of the increased transport channels. At the Fermi energy, there exist localized states confined in the CACs or in the inner side of the electrodes. Those localized states interact with the extending edge states and result in a sharp conductance peak at the Fermi energy.

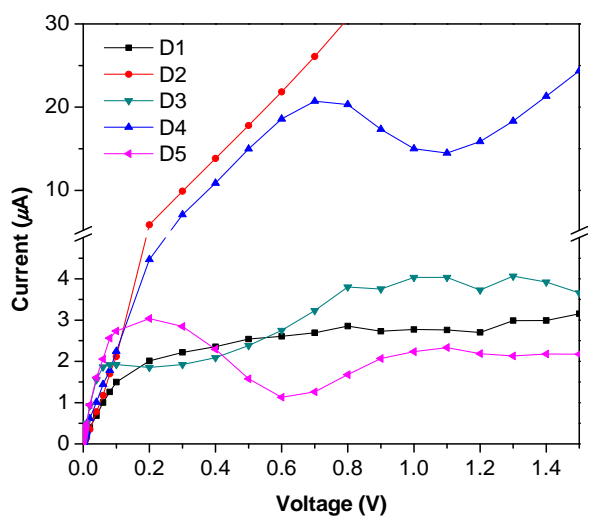

Figure 4. Currents versus the bias voltage are plotted for devices D1-D5.

In Figure 4, we plot the current-voltage curves of devices D1-D5. In the zero-bias linear transport region, all the devices show metallic behavior. Device D5 with triple CACs (purple) and device D3 with double side-CACs (blue) have the maxima conductance. This is a result of the fact that the electrons at the Fermi energy are in the edge states and the double side-CACs in D3 and D5 favorite their transport.

At finite bias voltage, however, the symmetry effects result in different behaviors of the devices. In the transport window, the energy range between the Fermi energies of the two electrodes, the conduction band with symmetric states in one electrode is aligned in energy to the valence band with antisymmetric states. The conductance spectra in the transport window become almost zero except near the boundaries in geometrically symmetric devices D1, D3, and D5. As a result, the current in those symmetric devices becomes limited as shown in Figure 4. The conductance of device D1 decreases gradually with the bias and the current saturates at bias higher than 0.5 V. This property suggests that the central-CAC geometry configuration may be used as a current stabilizer in circuits. The current in device D3 saturates quickly at 0.09 V after a linear increase before increasing very slowly again. In contrast, the two devices with asymmetric geometry, D2 and D4, show a linear I-V curve until much higher biases. When the two electrodes are connected by CACs in an asymmetric geometry, only part of the symmetric wave functions in one electrode can pass through the CACs and this part is not orthogonal to the asymmetric wavefunction in the other electrode. The conductance spectra can then be finite in the transport window. Device D2 (D4) remains an Ohmic contact up to a bias voltage of 1.5 V (0.6V). Finally, we observe the interesting NDR phenomenon in devices D4 and D5. The developing of the conductance spectra with the bias shows that the NDR is relevant to the conductance dip near energy 0.5eV in Figures 3e and 3g. The NDR occurs in the bias range [0.7, 1.1] V with a peak to valley ratio of 1.43 for device D4 and occurs in the range [0.2, 0.6] V with a ratio of 2.69 for device D5.

**Summary**

In summary, we have investigated the linear and nonlinear transport properties in junction devices of graphene nanoribbons connected by carbon atomic chains within the framework of density functional theory and the nonequilibrium Green's functions. The contact configuration plays a critical role in determining both the linear and the nonlinear properties. In the linear region, a square wave conductance spectrum can be realized where electrons above (below) the Fermi energy are allowed (forbidden) to pass through the device. In the nonlinear region, the negative differential resistance with a peak to valley ratio up to 2.69 can occur at low bias voltage in some cases and the Ohmic-contact or the stable-current connections can be realized in other cases. The ample diversity of transport properties in those systems indicates potential application of GNR-CAC junctions.

**Acknowledgment**

We thank Shuo-Wang Yang, Lei Shen, and P. Vasilopoulos for helpful discussion. This work was supported by the National Natural Science Foundation in China (Grant Nos. 11074182, 91121021, 11247276, and 11274238). It is partially supported by the Qing Lan Project and the project is funded by the Priority Academic Program Development of Jiangsu Higher Education Institutions.

**Table of Contents**

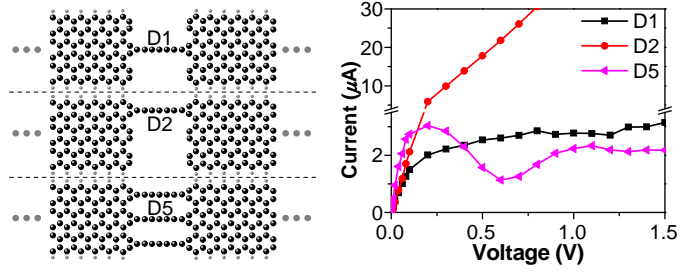
17**Table of Contents**

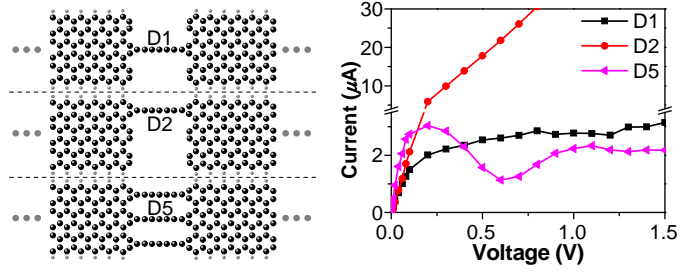